\documentstyle[aps,preprint,tighten]{revtex}

\newcommand{\beq}{\begin{eqnarray}}
\newcommand{\eeq}{\end{eqnarray}}

\newcommand{\la}{\langle}
\newcommand{\ra}{\rangle}
\newcommand{\q}{\la \bar{q}q \ra}
\newcommand{\uq}{\la \bar{u}u \ra}
\newcommand{\dq}{\la \bar{d}d \ra}
\newcommand{\s}{\la \bar{s}s \ra}
\newcommand{\gc}{\la {\alpha_s \over \pi} G^2 \ra}

\newcommand{\nc}{N_c \rightarrow \infty}

\begin{document}
\draft

\title{Multiquark picture for $\Sigma$ (1620)}
\author{Seungho Choe\thanks{Present address : Dept. of Physical Sciences, Hiroshima University,
Higashi-Hiroshima 739-8526, Japan (e-mail : schoe@hirohe.hepl.hiroshima-u.ac.jp).}}
\address{Special Research Centre for the Subatomic Structure of Matter, University
of Adelaide, Adelaide, SA 5005, Australia}
\maketitle
\begin{abstract}

In this work we report on a new QCD sum rule analysis to predict masses of
the excited baryon states (e.g. $\Sigma$ (1620) and $\Lambda$ (1405))
by using multiquark interpolating fields ($(q\bar{q})(qqq)$).
For the $\Sigma$ (1620) we consider the 
$\bar{K}N$, $\pi\Sigma$, and
$\pi\Lambda$ (I=1) multiquark interpolating fields.
The calculated mass from those multiquark states is about 1.592 GeV. 
For the $\Lambda$ (1405) we first show the result 
using the  
 $\pi^+\Sigma^- + \pi^0\Sigma^0 + \pi^-\Sigma^+$ (I=0) multiquark 
interpolating field, and compare the calculated mass to that
of our  
previous result using the 
$\pi^0\Sigma^0$ multiquark state.  
We then show that the mass 1.405 GeV is well reproduced when 
using the  $\bar{K}N$ (I=0) multiquark state.  
The uncertainties in our sum rules are also discussed.

\end{abstract}
\vspace{1cm}

\section{introduction}

QCD sum rule\cite{svz79,rry85,narison89} is a
 powerful tools to extract various properties of hadrons.
However, most QCD sum rule approaches have been applied to the lowest lying states,
and  it is rather difficult to properly extract physical properties of the
excited states.
For example, there have been only a limited number of 
 works on the excited baryons 
\cite{liu84,leinweber90,jko96,lk97,kl97} in which we are interested in this work.
 
Recently, we have proposed a new QCD sum rule analysis\cite{choe98} for calculating 
the  mass of
the $\Lambda$ (1405).  It was based on using the multiquark interpolating 
field ($(q\bar{q})(qqq)$) instead of the usual nucleon three quark
interpolating field $ (qqq)$.

In the case of the $\Lambda$ (1405) its nature is not revealed completely
yet, i.e. whether it is an ordinary three-quark state or a $\bar{K}N$
bound state or a mixed state of the previous two possibilities
\cite{pdg98}.
In Ref.\cite{choe98} we have focused on the decay channel of the
$\Lambda$ (1405) and introduced the $\pi^0\Sigma^0$ multiquark
interpolating field in order to get the $\Lambda$ (1405) mass
since the $\Lambda$ (1405) is only observed in the mass spectrum of the
$\pi\Sigma$ channel (I=0).
It has been found that the multiquark picture can be used
to extract physical properties of the excited baryons; e.g.
the mass of the excited baryon which is not fully accessible in the
conventional QCD sum rule approach.           

In this work we extend our previous analysis 
to the isospin I=1 multiquark states,
i.e. $\bar{K}N$, $\pi\Sigma$, and $\pi\Lambda$ multiquark states.
Our interpolating fields couple to both positive- and negative-parity, 
spin-${1\over 2}$ baryon states.
Among the $\Sigma$ particles (I=1), the lowest spin-${1\over 2}$
state  which couples to the $\bar{K}N$, $\pi\Sigma$, and $\pi\Lambda$ 
channels is the $\Sigma$ (1620) although the evidence of its existence is
only fair\cite{pdg98}.
We do not know the genuine structure of the $\Sigma$ (1620).
However, we can construct possible three multiquark states
 for the $\Sigma$ (1620) considering its decay channels.
Then, we can obtain the $\Sigma$ (1620) mass by following
the same procedures in Ref. \cite{choe98}, i.e. by comparing the mass
of the $\bar{K}N$, $\pi\Sigma$, and $\pi\Lambda$ multiquark states
each other.

In Sec.\ref{multi1} we present QCD sum rules for the I=1 multiquark states
and explain how to get the $\Sigma$ (1620) mass.
In Sec. \ref{multi0}
we also present a QCD sum rule for the $\Lambda$ (1405) mass
by taking into account the 
$\pi^+\Sigma^- + \pi^0\Sigma^0 + \pi^-\Sigma^+$ multiquark 
interpolating field, and compare it with the previous result for
the $\pi^0\Sigma^0$ multiquark interpolating field.
We discuss the uncertainties in our sum rules and summarize our results
in Sec.\ref{discuss}.

\section{QCD sum rules for I=1 multiquark states}{\label{multi1}}

Let's consider the following correlator:
\beq
\Pi (q^2) = i \int d^4x e^{iqx}\langle T ( J(x) \bar{J}(0) )\rangle ,
\eeq
where $J = \pi^+\Sigma^- - \pi^-\Sigma^+$ 
or $J =  \bar{K}^0 n - K^- p$, or $J = \pi\Lambda$
correspond 
to the multiquark interpolating fields for the isospin I=1 states.
The overall factor is irrelevant in our calculation.
Here, we take the interpolating fields for 
the nucleon, the $\Sigma$, and the $\Lambda$ particle as usual 
ones in the QCD sum rule calculations\cite{ioffe81,rry85}.
For example, in the case of the $\pi^0\Lambda$ multiquark state
we take $J = \epsilon_{abc}(\bar{u}_e i\gamma^5 u_e
			     -\bar{d}_e i\gamma^5 d_e)
([u_a^T C\gamma_\mu s_b]\gamma^5\gamma^\mu d_c
 - [d_a^T C\gamma_\mu s_b]\gamma^5\gamma^\mu u_c) $,
where $u$, $d$ and $s$ are the up, down and strange quark fields, and
$a,b,c,e$ are color indices.
$T$ denotes the transpose in Dirac space and $C$ is the charge
conjugation matrix.

The conventional QCD sum rule approach shows that the continuum effect
becomes larger with increasing the dimension of the interpolating field
and thus the results are more sensitive to a continuum
threshold.
However, as shown below we suggest a new approach to 
get the $\Sigma$ (1620) mass regardless of the large continuum effect.

In the case of the $\pi^+\Sigma^- - \pi^-\Sigma^+$
and $\bar{K}^0 n - K^- p$ multiquark interpolating fields
there are no exchange diagrams such as Fig. \ref{fig1}(a),
where the lowest two lines correspond to the quark fields of a meson
and the others are those of a baryon.
Then, for example, in the case of the $\bar{K}N$ multiquark states
the mass of the I=1 state is the same as
that of the I=0 state, i.e. the $\bar{K}^0 n + K^- p$ multiquark state.
Hence, for the $\bar{K}N$ and 
$\pi\Sigma$ (I=1) multiquark states
we use the $K^-p$ and $\pi^-\Sigma^+$ multiquark sum rules
in the previous work \cite{choe98}.

On the other hand, in the case of the $\pi\Lambda$ multiquark states
both the $\pi^0\Lambda$ and $\pi^{\pm}\Lambda$ multiquark 
interpolating fields give the same mass within
SU(2) symmetry (i.e. $m_u$ = $m_d$ = 0 and $\uq$ = $\dq$).
Thus, in what follows we present a QCD sum rule
for the $\pi^0\Lambda$ multiquark state only.
The OPE side has two structures:
\beq
\Pi^{OPE} (q^2) =\Pi_{q}^{OPE} (q^2) {\bf \rlap{/}{q}}
		+\Pi_{1}^{OPE} (q^2) {\bf 1} .
\eeq
In this paper, however, we only present the sum rule 
from the $\Pi_1$ structure (hereafter referred to as the $\Pi_1$ sum rule)
because the $\Pi_1$ sum rule (the chiral-odd sum rule)
 is generally more reliable than 
the $\Pi_q$ sum rule (the chiral-even sum rule)
 as emphasized in Ref. \cite{jt97}
and also in our previous work\cite{choe98}.
The OPE side is given as follows.
\beq
\Pi_{1}^{OPE} (q^2) = 
 &+& \frac{11 ~m_s}{\pi^8 ~2^{18} ~3^2 ~5^2} q^{10} ln(-q^2)
+ \frac{1}{\pi^6 ~2^{15} ~3^2 ~5} (40\q - 11\s) q^8 ln(-q^2)
\nonumber\\
&-& \frac{m_s^2}{\pi^6 ~2^{14} ~3^2} (80\q + 11\s) q^6 ln(-q^2)
\nonumber\\
&-& \frac{m_s}{\pi^4 ~2^9 ~3^2} (23\q^2 - 20\q \s) q^4 ln (-q^2)
\nonumber\\
&+& \frac{1}{\pi^2 ~2^6 ~3^2} (40\q^3 + 45\q^2 \s) q^2 ln(-q^2)
\nonumber\\
&-& \frac{m_s^2}{\pi^2 ~2^6 ~3^2} (26\q^3 - 45\q^2\s) ln(-q^2)
\nonumber\\
&-& \frac{m_s}{2^4 ~3^3} (132\q^4 - 37\q^3 \s) \frac{1}{q^2} ,
\label{ope_pilam}
\eeq
where $m_s$ is the strange quark mass and $\q$, $\s$ are the 
quark condensate and the strange quark condensate, respectively.
Here, we let
$m_u$ = $m_d$ = 0 $\neq$ $m_s$ and
$\uq$ = $\dq$ $\equiv$ $\q$ $\neq$ $\s$.
We neglect the contribution of gluon condensates and
concentrate on tree diagrams such as
Figs. \ref{fig1}(a) and \ref{fig1}(b)
(hereafter referred to as ``bound" diagrams and ``unbound" diagrams,
respectively),
and assume the vacuum saturation hypothesis
to calculate quark condensates of higher dimensions.
Note that only some typical diagrams are shown in Fig. \ref{fig1}.

The contribution of the ``bound'' diagrams is a $1/N_c$ correction
to that of the ``unbound'' diagrams, 
where $N_c$ is the number of the colors.
In Eq. (\ref{ope_pilam}) and in what follows we set $N_c$ = 3.
The ``unbound'' diagrams correspond to a picture that two particles
are flying away without any interaction between them.
In the $\nc$ limit
only the ``unbound'' diagrams contribute to the $\pi\Lambda$
multiquark sum rule.
Then, the $\pi\Lambda$ multiquark mass should be
the sum of the pion and the $\Lambda$ mass in this limit.
For the sake of reference, 
we present the OPE side in the $\nc$ limit in the below.
\beq
\Pi_{1}^{OPE(\nc)}(q^2) = 
 &+& \frac{m_s}{\pi^8 ~2^{16} ~3 ~5^2} q^{10} ln(-q^2)
+ \frac{1}{\pi^6 ~2^{13} ~3 ~5} (4\q - \s) q^8 ln(-q^2)
\nonumber\\
&-& \frac{m_s^2}{\pi^6 ~2^{12} ~3} (8\q + \s) q^6 ln(-q^2)
\nonumber\\
&-& \frac{m_s}{\pi^4 ~2^8 ~3^2} (19\q^2 - 12\q \s) q^4 ln (-q^2)
\nonumber\\
&+& \frac{1}{\pi^2 ~2^4 ~3} (4\q^3 + 5\q^2 \s) q^2 ln(-q^2)
\nonumber\\
&-& \frac{m_s^2}{\pi^2 ~2^4 ~3} (4\q^3 - 5\q^2\s) ln(-q^2)
\nonumber\\
&-& \frac{m_s}{3^2} (3\q^4 - \q^3 \s) \frac{1}{q^2} .
\label{ope_pilam_nc}
\eeq
The OPE sides in Eqs. (\ref{ope_pilam}) and (\ref{ope_pilam_nc}) 
have the following form:
\beq
\Pi^{OPE}_1 (q^2) &=& a ~q^{10} ln(-q^2) + b ~q^8 ln(-q^2)
+ c ~q^6 ln(-q^2) 
+ d ~q^4 ln(-q^2)
\nonumber \\
&+& e ~q^2 ln(-q^2)+ f ~ln(-q^2) + g ~\frac{1}{q^2} ,
\eeq
where $a, b, c, \cdots, g$ are constants. Then we parameterize the
phenomenological side as
\beq
\frac{1}{\pi} Im \Pi^{Phen}_{1} (s) &=& \lambda^2 m \delta(s-m^2) +
	     [-a~s^5 - b~s^4 - c~s^3 - d~s^2 - e~s - f] \theta(s~-~s_0) ,
\eeq
where $m$ is the $\pi\Lambda$ multiquark mass
and $s_0$ the continuum threshold.
$\lambda$ is the coupling strength of the interpolating field
to the physical $\Sigma$ (1620) state.
After Borel transformation
the mass $m$ is given by
\beq
m^2 &=& M^2 \times
\nonumber\\
&\{&-720a (1-\Sigma_6)
- \frac{120b}{M^2} (1-\Sigma_5)
- \frac{24c}{M^4} (1-\Sigma_4)
\nonumber\\
&-& \frac{6d}{M^6} (1-\Sigma_3)
- \frac{2e}{M^8} (1-\Sigma_2)
- \frac{f}{M^{10}} (1-\Sigma_1) \}
~/
\nonumber\\
&\{&-120a (1-\Sigma_5)
- \frac{24b}{M^2} (1-\Sigma_4)
- \frac{6c}{M^4} (1-\Sigma_3)
\nonumber\\
&-& \frac{2d}{M^6} (1-\Sigma_2)
-\frac{e}{M^8} (1-\Sigma_1)
- \frac{f}{M^{10}} (1-\Sigma_0)
- \frac{g}{M^{12}} \} ,
\label{mass}
\eeq
where
\beq
\Sigma_i = \sum_{k=0}^i \frac{s_0^k}{k~! ~(M^2)^k} ~e^{-\frac{s_0}{M^2}}  .
\eeq
Fig. \ref{fig2} shows the Borel-mass dependence of the $\pi\Lambda$
multiquark mass at $s_0$ = 2.756 GeV$^2$ taken by considering the next 
$\Sigma$ (1660) \cite{pdg98}.
There is a plateau for the large Borel mass, but this is a trivial result 
from our crude model on the phenomenological side.
Hence we do not take this as the $\pi\Lambda$ multiquark mass
and neither as the $\Sigma$ (1620) mass.

Instead, we draw the Borel-mass
dependence of the coupling strength
$\lambda^2$ at $s_0$ = 2.756 GeV$^2$ in Fig. \ref{fig3}.
There is the maximum point in the figure.
It means that the $\pi\Lambda$ multiquark interpolating fields couples
strongly to the physical $\Sigma$ (1620) state at this point.
Then we take the $\Sigma$ (1620) mass as 
that of the $\pi\Lambda$ multiquark state at the point.
However, $s_0$ is taken by hand considering the experimental
$\Sigma$ (1660) mass.
It would be better to determine an effective threshold $s_0$ from 
the present sum rule itself. In what follows we explain how to determine
the effective threshold and thus the $\Sigma$ (1620) mass.

The $\Sigma$ (1620) is the lowest spin-$1\over2$ state which couples to
the three I=1 multiquark states, i.e.
the $\bar{K}N$, $\pi\Sigma$, and $\pi\Lambda$ states.
The aim of the present work is to get the masses of the 
$\bar{K}N$, $\pi\Sigma$, and $\pi\Lambda$ multiquark states,
where the coupling strength of each multiquark state has its maximum value.
We take the same threshold for each multiquark sum rule.
Why we can compare the multiquark masses at the same threshold
although they are obtained with different interpolating fields
will be clear later.
We choose the threshold in order that
the $\bar{K}N$ multiquark mass
becomes the sum of the kaon and the nucleon mass at least.
Then, above the threshold the $\Sigma$ (I=1) particle can couple
to the $\pi\Sigma$, $\pi\Lambda$ and $\bar{K}N$ multiquark states,
while below the threshold
to the $\pi\Sigma$ and/or $\pi\Lambda$ multiquark state(s) only.

Let us consider the $\bar{K}N$ multiquark sum rule to get the effective
threshold.
The OPE side is the same as that of
the $K^+p$ multiquark sum rule without 
the contribution of ``bound'' diagrams\cite{choe98}, and thus given by
\beq
\Pi^{OPE(\nc)}_1 (q^2) = 
 &+& \frac{1}{\pi^6 ~2^{13} ~3 ~5} \q q^8 ln(-q^2)
 - \frac{m_s^2}{\pi^6 ~2^{11} ~3} \q q^6 ln(-q^2)
\nonumber\\
&-& \frac{m_s}{\pi^4 ~2^8 ~3} (2\q^2 - \q \s) q^4 ln (-q^2)
\nonumber\\
&+& \frac{1}{\pi^2 ~2^4 ~3} (2\q^3 + \q^2 \s) q^2 ln(-q^2)
\nonumber\\
&-& \frac{m_s^2}{\pi^2 ~2^4 ~3} (4\q^3 - \q^2\s) ln(-q^2)
\nonumber\\
&-& \frac{m_s}{2 ~3^2} (2\q^4 - \q^3 \s) \frac{1}{q^2} .
\label{ope_kn}
\eeq
Fig. \ref{fig4} presents the Borel-mass dependence of the
$\bar{K}N$ multiquark mass in the fiducial Borel interval
which lies in the 30 \% -- 50 \% criteria,
i.e. the contribution of the power correction
is less than 30 \% and that of the continuum is less than 50 \%.
We define 
$A \equiv M^2 \times {- 120 b \over - 24 b}$ and
$B \equiv
M^2 \times {-120b - \frac{24c}{M^2} - \frac{6d}{M^4}
 - \frac{2e}{M^6} - \frac{f}{M^8}
\over
-24b - \frac{6c}{M^2} - \frac{2d}{M^4} - \frac{e}{M^6}
 -\frac{f}{M^8} - \frac{g}{M^{10}}}$.
Then, we calculate the contribution of the power correction as
$C \equiv 1-{\sqrt{A} \over \sqrt{B}}$.
Using this factor we determine the lower bound of the Borel interval
where $C$ is 0.3 at most. 
On the other hand, we choose the upper bound of the Borel interval
in order that the factor
$D \equiv 1-{\sqrt{Eq. (\ref{mass})} \over \sqrt{B}}$
is 0.5 at most.
Since there is no plateau in the fiducial Borel interval in
Fig. \ref{fig4},
we take the $\bar{K}N$ multiquark mass as an average value
in the interval.
Then, we get $s_0$ = 3.852 GeV$^2$ where the average mass
becomes $m_K$ + $m_p$ = 1.435 GeV.

Now, let us go back to the $\Sigma$ (1620) mass.
We redraw the Borel-mass dependence of the coupling strength
from the $\pi\Lambda$
multiquark sum rule at $s_0$ = 3.852 GeV$^2$ obtained in the above, and then
take the $\Sigma$ (1620) mass as the value 
where the coupling strength
has its maximum value.
In Table \ref{mass_i1} we present the mass $m$ for each
multiquark state, where we take $\q$ = -- (0.230 GeV)$^3$, 
$\s$ = 0.8~$\q$, and $m_s$ = 0.150 GeV as input parameters.
In the case of the $\bar{K}N$ and $\pi\Sigma$
multiquark states we use the same masses in Ref. \cite{choe98}
obtained from the $K^-p$ and the $\pi^-\Sigma^+$ multiquark state, 
respectively.
Fig. \ref{fig5} shows the coupling strength and the mass from each
multiquark state at $s_0$ = 3.852 GeV$^2$.
The average mass from three states in the table is about 1.592 GeV,
and it is rather smaller than the experimental
value, 1.620 GeV\cite{pdg98}.
Fig. \ref{fig6} presents the Borel-mass dependence of
the $\pi\Lambda$ multiquark mass on the strange quark mass, 
the strange quark condensate, and the quark condensate, respectively,
at $s_0$ = 3.852 GeV$^2$.
It seems that the SU(3) symmetry breaking effects are not significant
in our sum rule.

It is interesting to note that the masses in Table \ref{mass_i1}
are very similar.
We have checked that at an arbitrary threshold 
three multiquark sum rules give similar masses.
Thus, in principle we can predict another mass of the $\Sigma$ 
excited state using
these three multiquark sum rules if the threshold is taken properly.

Let us stop here to remark why the multiquark masses are similar 
although they are calculated from different interpolating fields,
i.e. the $\bar{K}N$, $\pi\Sigma$ and $ \pi\Lambda$ multiquark
interpolating fields, respectively.
First of all, comparing Eq. (\ref{ope_pilam_nc}) and Eq. (\ref{ope_kn})
one can easily find that they have the same structure within SU(3) symmetry
($m_u = m_d = m_s = 0$, $\q = \s$).
For completeness, we write down the OPE side for the $\pi\Sigma$
multiquark interpolating field in the $\nc$ limit.
\beq
\Pi_{1}^{OPE(\nc)} (q^2) = 
 &-& \frac{m_s}{\pi^8 ~2^{16} ~3 ~5^2} q^{10} ln(-q^2)
+ \frac{1}{\pi^6 ~2^{13} ~3 ~5} \s q^8 ln(-q^2)
\nonumber\\
&+& \frac{m_s^2}{\pi^6 ~2^{12} ~3} \s q^6 ln(-q^2)
- \frac{m_s}{\pi^4 ~2^7} \q^2 q^4 ln (-q^2)
\nonumber\\
&+& \frac{1}{\pi^2 ~2^4} \q^2 \s q^2 ln(-q^2)
+ \frac{m_s^2}{\pi^2 ~2^4} \q^2\s ln(-q^2)
\nonumber\\
&-& \frac{m_s}{3^2} \q^4 \frac{1}{q^2} .
\label{ope_pisig_i1}
\eeq
Of course, as we said before, there is no contribution of the ``bound" diagrams
for the $\bar{K}N$ and $\pi\Sigma$ multiquark sum rules.
Second, as shown in Fig. \ref{fig6} the SU(3) symmetry breaking
effects are small in our sum rules. In Fig. \ref{fig7} we plot the
$\pi\Lambda$ multiquark mass with(the solid line) and without(the dotted line)
the SU(3) symmetry breaking effects.
We draw the solid line using 
$\q$ = -- (0.230 GeV)$^3$, $\s$ = 0.8~$\q$, and $m_s$ = 0.150 GeV,
while draw the dotted line taking
$\q$ = -- (0.230 GeV)$^3$, $\s$ = 1.0~$\q$, and $m_s$ = 0.
Last, the contribution of the ``bound" diagrams in the $\pi\Lambda$
multiquark sum rule is very small as shown
in Fig. \ref{fig8}.
The solid line is the $\pi\Lambda$ multiquark mass from 
all diagrams; i.e. the ``unbound" + ``bound" diagrams
(from Eq. (\ref{ope_pilam})), while the dotted line is
the mass from the ``unbound" diagrams only (from Eq. (\ref{ope_pilam_nc})). 
Because of the above reasons 
the masses from three multiquark sum rules are similar.
Hence, we can apply the same threshold for each multiquark sum rule
and compare the multiquark mass each other.

\section{QCD sum rules for I=0 multiquark states}{\label{multi0}}

In the following we get the  $\Lambda$ (1405) mass
by taking into account the 
 $\pi^+\Sigma^- + \pi^0\Sigma^0 + \pi^-\Sigma^+$ (I=0) multiquark 
interpolating field and compare the mass with
 the previous result for the
$\pi^0\Sigma^0$ multiquark state\cite{choe98}.
The $\pi^+\Sigma^- + \pi^0\Sigma^0 + \pi^-\Sigma^+$ multiquark state
is the complete basis for the I=0 state in contrast to the $\pi^0\Sigma^0$
multiquark state. 

The $\Pi^{OPE}_1$ for the
 $\pi^+\Sigma^- + \pi^0\Sigma^0 + \pi^-\Sigma^+$ multiquark
state is given by 
\beq
\Pi_{1}^{OPE} (q^2) = 
 &-& \frac{7 ~m_s}{\pi^8 ~2^{18} ~3^2 ~5} q^{10} ln(-q^2)
+ \frac{7}{\pi^6 ~2^{15} ~3^2} \s q^8 ln(-q^2)
\nonumber\\
&+& \frac{35 ~m_s^2}{\pi^6 ~2^{14} ~3^2} \s q^6 ln(-q^2)
- \frac{121 ~m_s}{\pi^4 ~2^9 ~3^2} \q^2 q^4 ln (-q^2)
\nonumber\\
&+& \frac{11}{\pi^2 ~2^6} \q^2 \s q^2 ln(-q^2)
- \frac{m_s^2}{\pi^2 ~2^6 ~3} (14\q^3 - 33 \q^2\s) ln(-q^2)
\nonumber\\
&-& \frac{m_s}{2^4 ~3^3} (140\q^4 + 3\q^3 \s) \frac{1}{q^2} ,
\label{ope_pisig_i0}
\eeq
and in the $\nc$ limit 
\beq
\Pi_{1}^{OPE(\nc)} (q^2) = 
 &-& \frac{m_s}{\pi^8 ~2^{16} ~5^2} q^{10} ln(-q^2)
+ \frac{1}{\pi^6 ~2^{13} ~5} \s q^8 ln(-q^2)
\nonumber\\
&+& \frac{m_s^2}{\pi^6 ~2^{12}} \s q^6 ln(-q^2)
- \frac{65 ~m_s}{\pi^4 ~2^8 ~3^2} \q^2 q^4 ln (-q^2)
\nonumber\\
&+& \frac{3}{\pi^2 ~2^4} \q^2 \s q^2 ln(-q^2)
- \frac{m_s^2}{\pi^2 ~2^4 ~3} (4\q^3 - 9\q^2\s) ln(-q^2)
\nonumber\\
&-& \frac{m_s}{3} \q^4 \frac{1}{q^2} .  
\label{ope_pisig_i0_nc}
\eeq
In the case of the $\Lambda$ (1405) it couples to the $\pi\Sigma$
channel only. 
Thus, we take the threshold $s_0$ = 3.082 GeV$^2$
in order that the $\pi\Sigma$ (I=0) multiquark mass
becomes the sum of the pion and the $\Sigma$ mass
in the fiducial Borel interval when only the ``unbound" diagrams are considered.
Here, we use the average mass of the pions and the $\Sigma$ particles,
i.e. 0.138 + 1.193 = 1.331 GeV.
In order to get the Borel interval  we define new parameters 
$C' \equiv 1-{\sqrt{A'} \over \sqrt{B'}}$ and
$D' \equiv 1-{\sqrt{Eq. (\ref{mass})} \over \sqrt{B'}}$, where
$A' \equiv M^2 \times {- 720 a \over - 120 a}$ and
$B' \equiv
M^2 \times {-720a - \frac{120b}{M^2} - \frac{24c}{M^4}
 - \frac{6d}{M^6} - \frac{2e}{M^8}
 - \frac{f}{M^{10}}
\over
-120a - \frac{24b}{M^2} - \frac{6c}{M^4} - \frac{2d}{M^6}
 -\frac{e}{M^8} - \frac{f}{M^{10}} - \frac{g}{M^{12}}}$.
Following the same procedures in the previous section
we get the $\Lambda$ (1405) mass as shown in Table \ref{mass_i0qc}.
The mass is very similar to that of
the $\pi^0\Sigma^0$ multiquark state.
In Table \ref{mass_i0qc} we also present the variation of 
the $\Lambda$ (1405) mass on the quark condensate. The mass becomes
smaller as the absolute value of the quark condensate increases.

On the other hand, the $\bar{K}N$ (I=0) multiquark mass 
at $s_0$ = 3.082 GeV$^2$ is 1.405 GeV,
and it is closer  
to the experimental value comparing to 1.387 GeV which was 
obtained previously
using the threshold
$s_0$ = 3.012 GeV$^2$ from the $\pi^0\Sigma^0$ multiquark sum rule\cite{choe98}.
Fig. \ref{fig9} shows the coupling strength and the mass from the $\bar{K}N$
and $\pi^+\Sigma^- + \pi^0\Sigma^0 + \pi^-\Sigma^+$ multiquark state, respectively.
 
One can easily find that within SU(3) symmetry ($m_u = m_d = m_s =0$, $\q = \s$)
Eq. (\ref{ope_pisig_i0_nc}) has the same structure as in Eq. (\ref{ope_kn}).
Note that the $\bar{K}N$ (I=0) multiquark sum rule is the same
as the $\bar{K}N$ (I=1) multiquark sum rule since there is no
contribution of the ``bound" diagrams for the $\bar{K}N$ (both I=0 and I=1) multiquark 
interpolating fields.

Table \ref{mass_i0} shows the masses from the
I=0 multiquark states at $s_0$ = 3.852 GeV$^2$,
and these values correspond to the $\Lambda$ (1600) mass.
Because the $\Lambda$ (1600) couples to both
the $\bar{K}N$ and $\pi\Sigma$ channels (I=0), 
we get the $\Lambda$ (1600) mass by comparing the $\bar{K}N$
 and $\pi\Sigma$ multiquark mass
at the same threshold.
The average mass between the $\bar{K}N$ and
 $\pi^+\Sigma^- + \pi^0\Sigma^0 + \pi^-\Sigma^+$ (I=0) multiquark 
states in the table is 1.601 GeV.

\section{discussion}{\label{discuss}}

Let us discuss the uncertainties in our results.
Comparing Tables \ref{mass_i1} and \ref{mass_i0}, 
each average mass of the I=0 and I=1 multiquark states is
slightly different from the experimental values,
i.e. the $\Lambda$ (1600) and the $\Sigma$ (1620), respectively.
Note that in the previous sections we have used the same threshold
for the I=1 and I=0 $\bar{K}N$ multiquark states because
we can not distinguish the I=1 state from the
I=0 state in our approach.

The mass difference 
can be calculated by
including the isospin symmetry breaking effects (i.e. $m_u \neq m_d \neq 0$,
$\uq \neq \dq$, and electromagnetic effects) in our sum rules 
as in Refs. \cite{hhp90,yhhk93,ei93,adi93,hy94,jin95,kl95,zl97}. 
If this correction is included,
then the threshold for the $\bar{K}N$ multiquark state 
will be different from the previous one.
Then, the masses of other multiquark states
(both the I=0 and I=1 states)
can also be changed according to the new threshold.

On the other hand, one can consider the contractions between 
the $\bar{u}$ and $u$ (or between the $\bar{d}$ and $d$) quarks
in the initial state which have been excluded in our previous calculation.
If this correction is taken into account, then we can
distinguish between the $\bar{K}^0 n + K^- p$ (I=0) and
$\bar{K}^0 n - K^- p$ (I=1) multiquark states 
because it is one of $1/N_c$ corrections.
Although we are only interested in the five-quark states 
in the initial state,
one can check easily the amount of contribution to the
previous calculation.
Including this correction 
the mass for the  $\bar{K}^0 n + K^- p$ (I=0) multiquark state
becomes 1.590 GeV while 1.588 GeV 
for the $\bar{K}^0 n - K^- p$ (I=1) multiquark state
at $s_0$ = 3.852 GeV$^2$.
Note that the effective threshold $s_0$ is obtained by including
the ``unbound'' diagrams only and thus we can use the same threshold 
for both multiquark states.
The masses of other multiquark states are rarely 
changed even if this correction is considered.
It is found that this correction
is very small comparing to other $1/N_c$ corrections,
i.e. the contribution of ``bound" diagrams.
Another possibility  to get the mass difference will be the correction from
the possible instanton effects\cite{ss98,dk90,fb93} to the I=0 and
I=1 states, respectively.    

In this work we have neglected the contribution of gluon condensates.
Since we have considered the $\Pi_1$ sum rule (the chiral-odd sum rule),
 only the odd
dimensional operators can contribute to the sum rule.
For example, the contribution of the gluon condensates
is given by the terms like
$m_s \gc $ and thus can be neglected
comparing to other quark condensates of the same dimension.
However, as shown in the nucleon sum rule,
the gluon condensate term significantly
affects the Borel-stability although it is numerically small.
In this respect, further analysis including the contribution of
the gluon condensates in 
our sum rules is required before any firm conclusions may be drawn.

In summary, the $\Sigma$ (1620) mass is predicted in the
QCD sum rule approach using
the $\bar{K}N$, $\pi\Sigma$, and $\pi\Lambda$ (I=1)
multiquark interpolating fields.
The mass from the $\Pi_1$ sum rule
(the chiral-odd sum rule) is about 1.592 GeV.
The $\Lambda$ (1405) mass is also obtained considering
the $\pi^+\Sigma^- + \pi^0\Sigma^0 + \pi^-\Sigma^+$ (I=0) 
multiquark state. The mass is 1.424 GeV,  while that of the $\bar{K}N$ (I=0)
multiquark state is 1.405 GeV.
On the other hand, it would be interesting to calculate 
the $\Sigma$ (1620) mass by following the methods in Refs.\cite{lk97,kl97}
which have obtained the N (1535) and the $\Lambda$ (1405)
mass, respectively, using the interpolating fields with 
a covariant derivative. 

\acknowledgements

The author thanks Prof. A.W. Thomas, Prof. Su H. Lee, and Prof. O. Miyamura 
for valuable comments.
This work is supported by Research Fellowship of the Japan Society for the
Promotion of Science (JSPS), and also supported in part by
Centre for the Subatomic Structure of Matter (CSSM)
at Adelaide University where it was started.



\begin{table}[t]
\caption{Mass of the $\bar{K}N$, $\pi\Sigma$, and $\pi\Lambda$ (I=1)
multiquark states at $s_0$ = 3.852 GeV$^2$
(~$\q$ = -- (0.230 GeV)$^3$, $\s$ = 0.8~$\q$, and $m_s$ = 0.150 GeV). }
\label{mass_i1}
\begin{center}
\begin{tabular}{c c c c}
  &  multiquark state      &  $m$(GeV)  & \\
\hline
  &  $\bar{K}N$            &  1.589 & \\
  &  $\pi\Sigma$           &  1.606 & \\
  &  $\pi\Lambda$          &  1.581 & \\
\end{tabular}
\end{center}
\end{table}

\begin{table}[t]
\caption{Mass of  the $\pi^+\Sigma^- + \pi^0\Sigma^0 + \pi^-\Sigma^+$ (I=0)
multiquark state (~$\s$ = 0.8~$\q$, $m_s$ = 0.150 GeV).
[$\cdots$] means the value from the $\pi^0\Sigma^0$ multiquark state. }
\label{mass_i0qc}
\begin{center}
\begin{tabular}{c c c c c}
 & quark condensate (GeV$^3$)  & $s_0$ (GeV$^2$)  &  $m$(GeV) & \\
\hline
 &  --(0.210)$^3$ &  3.093 [3.015]  &  1.443 [1.434] & \\
 &  --(0.230)$^3$ &  3.082 [3.012]  &  1.424 [1.419] & \\
 &  --(0.250)$^3$ &  3.077 [3.008]  &  1.409 [1.404] & \\
\end{tabular}
\end{center}
\end{table}

\begin{table}[t]
\caption{Mass of the $\bar{K}N$ and $\pi\Sigma$ (I=0)
multiquark states at $s_0$ = 3.852 GeV$^2$
(~$\q$ = -- (0.230 GeV)$^3$, $\s$ = 0.8~$\q$, and $m_s$ = 0.150 GeV). 
[$\cdots$] means the value from the $\pi^0\Sigma^0$ multiquark state. }
\label{mass_i0}
\begin{center}
\begin{tabular}{c c c c}
  & multiquark state      &  $m$(GeV) & \\
\hline
  & $\bar{K}N$            &  1.589 & \\
  & $\pi^+\Sigma^- + \pi^0\Sigma^0 + \pi^-\Sigma^+$
                          &  1.612 [1.625] & \\
\end{tabular}
\end{center}
\end{table}


\begin{figure}
\caption{Diagrams. Solid lines are the quark propagators. 
(a) ``bound'' diagrams (b) ``unbound''  diagrams.}
\label{fig1}
\end{figure}

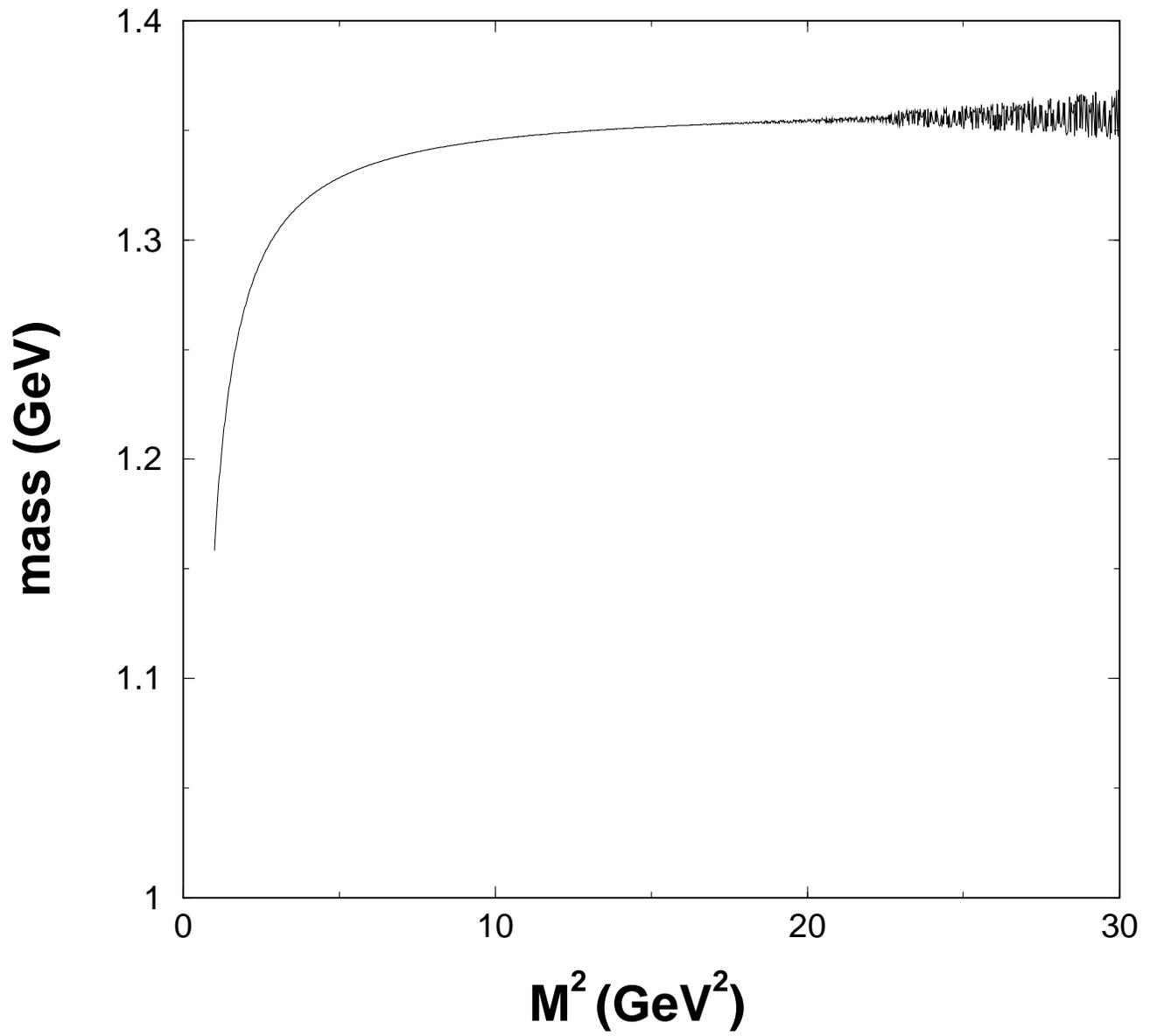
\begin{figure}
\caption{The Borel-mass dependence of the 
$\pi\Lambda$ multiquark mass at $s_0$ = 2.756 GeV$^2$.}
\label{fig2}
\end{figure}

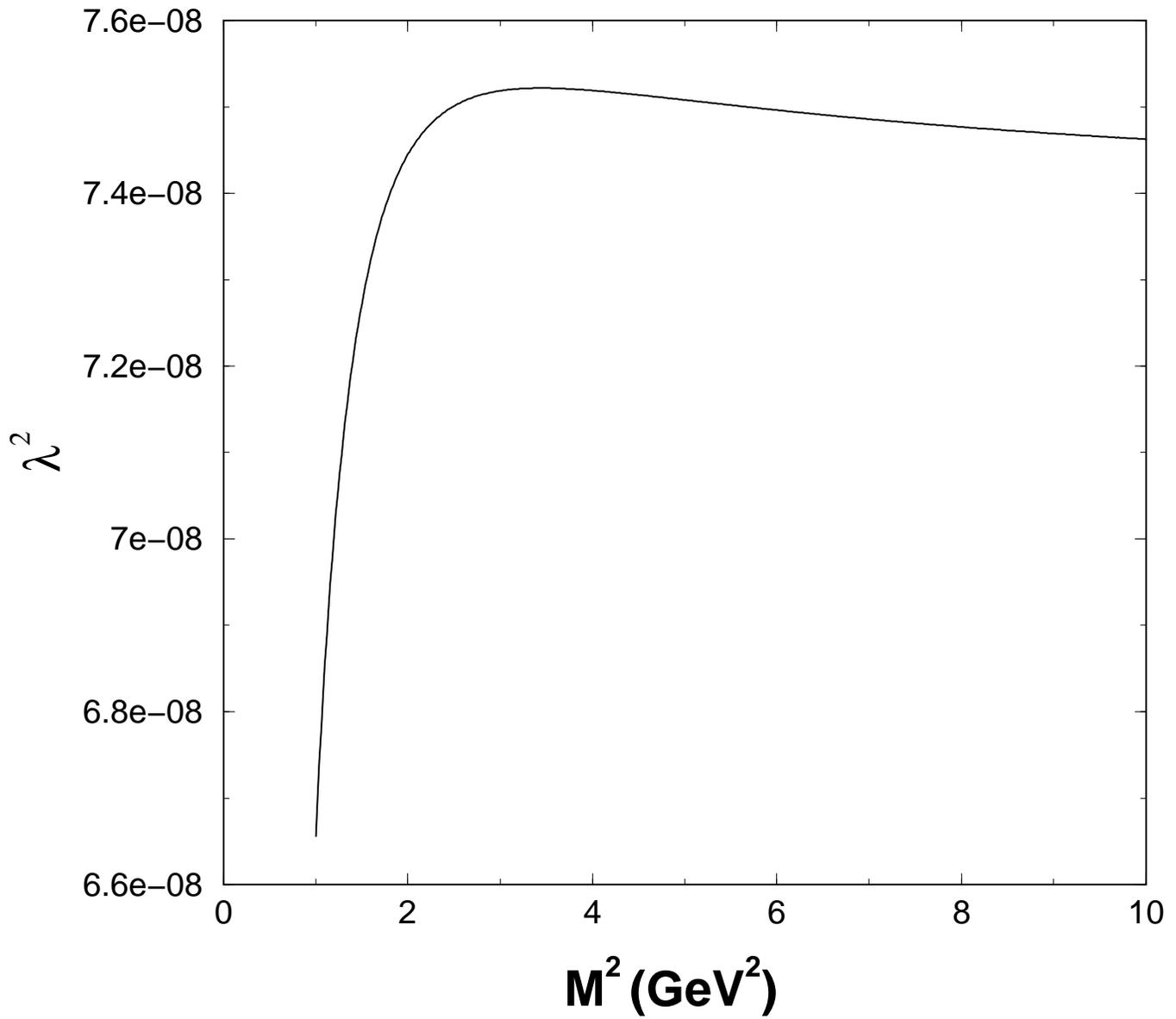
\begin{figure}
\caption{The Borel-mass dependence of the coupling strength 
$\lambda^2$ from the $\pi\Lambda$ multiquark sum rule at 
$s_0$ = 2.756 GeV$^2$.}
\label{fig3}
\end{figure}

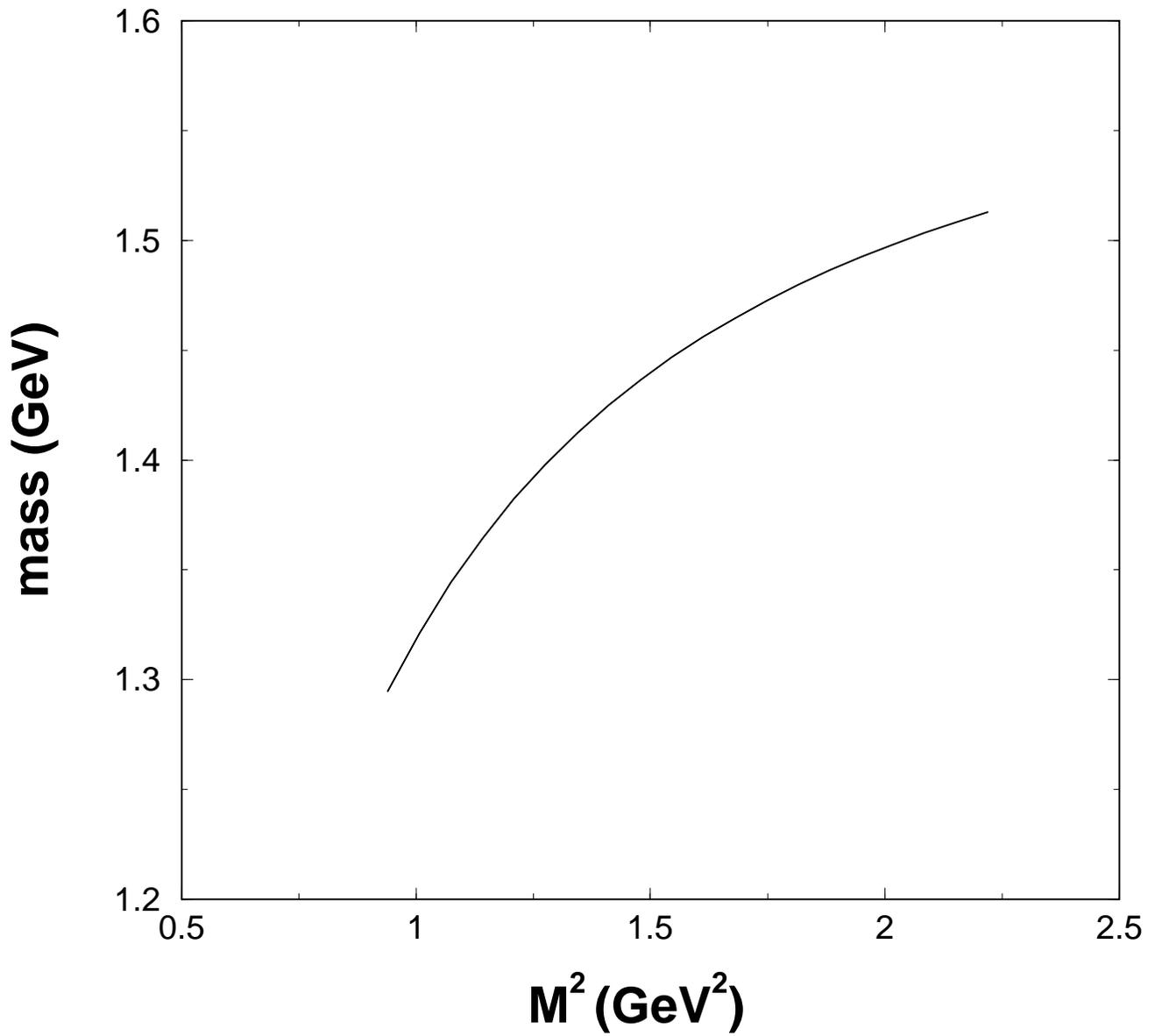
\begin{figure}
\caption{The Borel-mass dependence of the $\bar{K}N$ multiquark 
mass in the fiducial Borel interval at $s_0$ = 3.852 GeV$^2$.}
\label{fig4}
\end{figure}

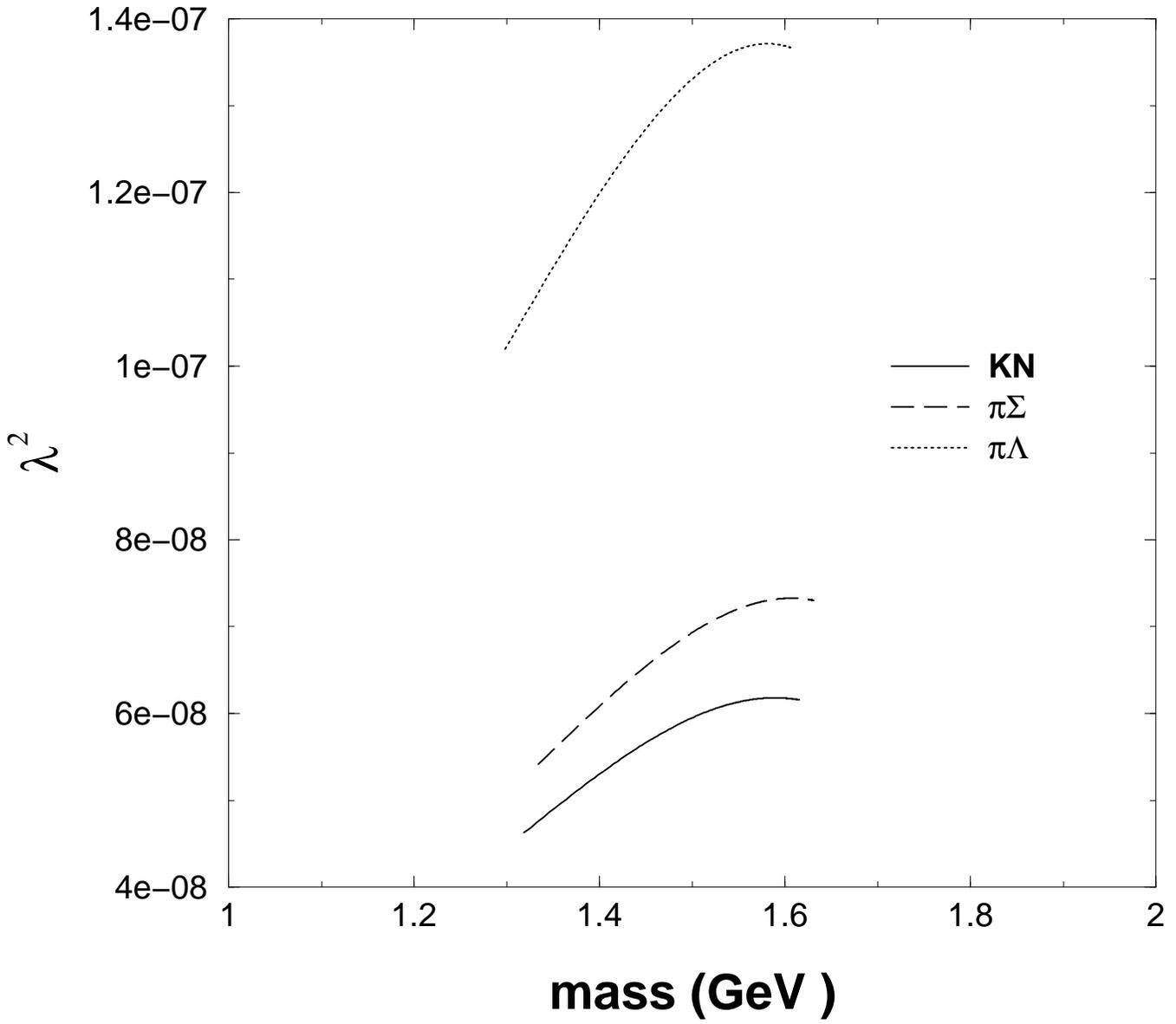
\begin{figure}
\caption{The coupling strengths and masses from the $\bar{K}N$,
$\pi\Sigma$, and $\pi\Lambda$ multiquark states
at $s_0$ = 3.852 GeV$^2$.}
\label{fig5}
\end{figure}

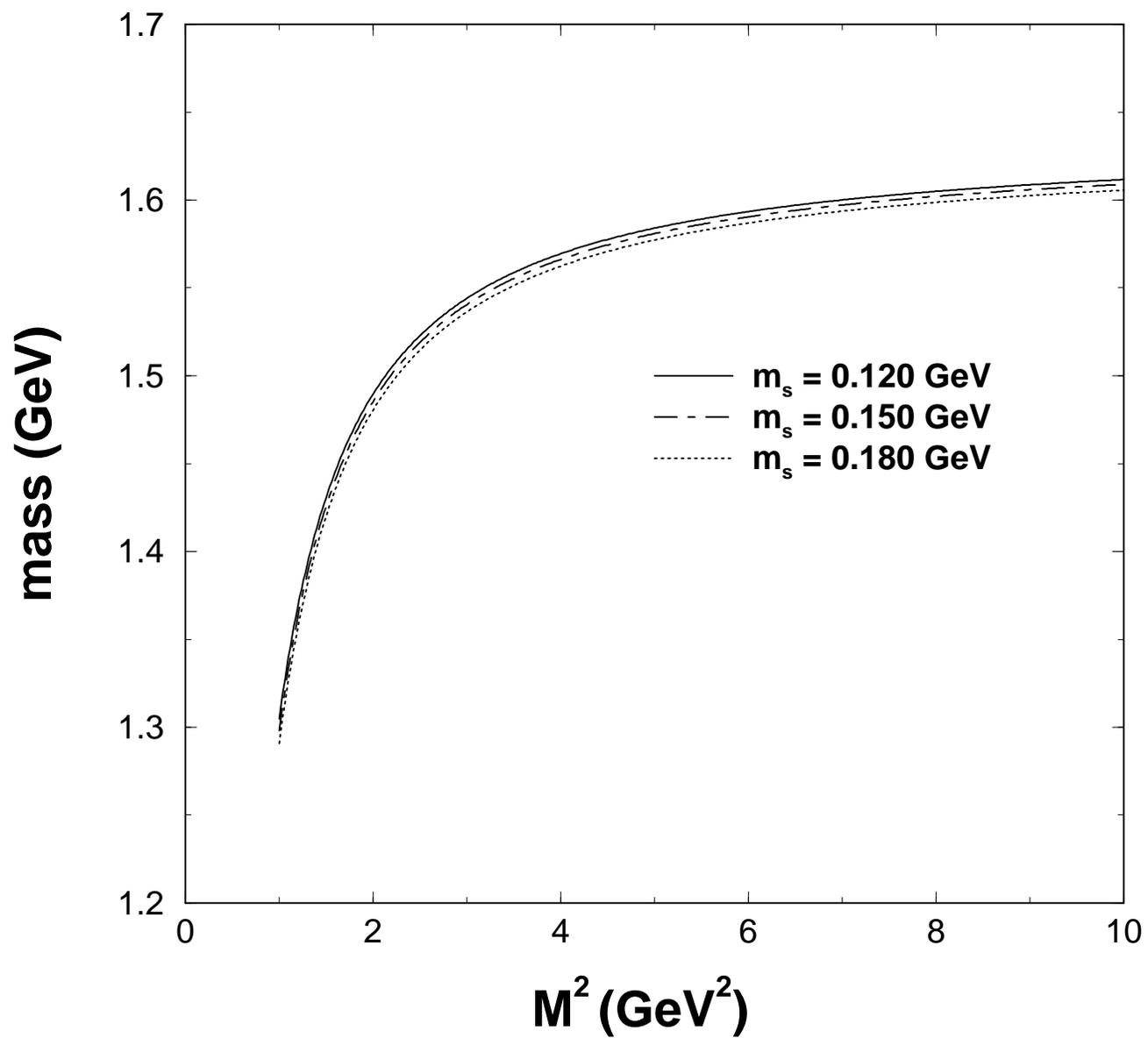
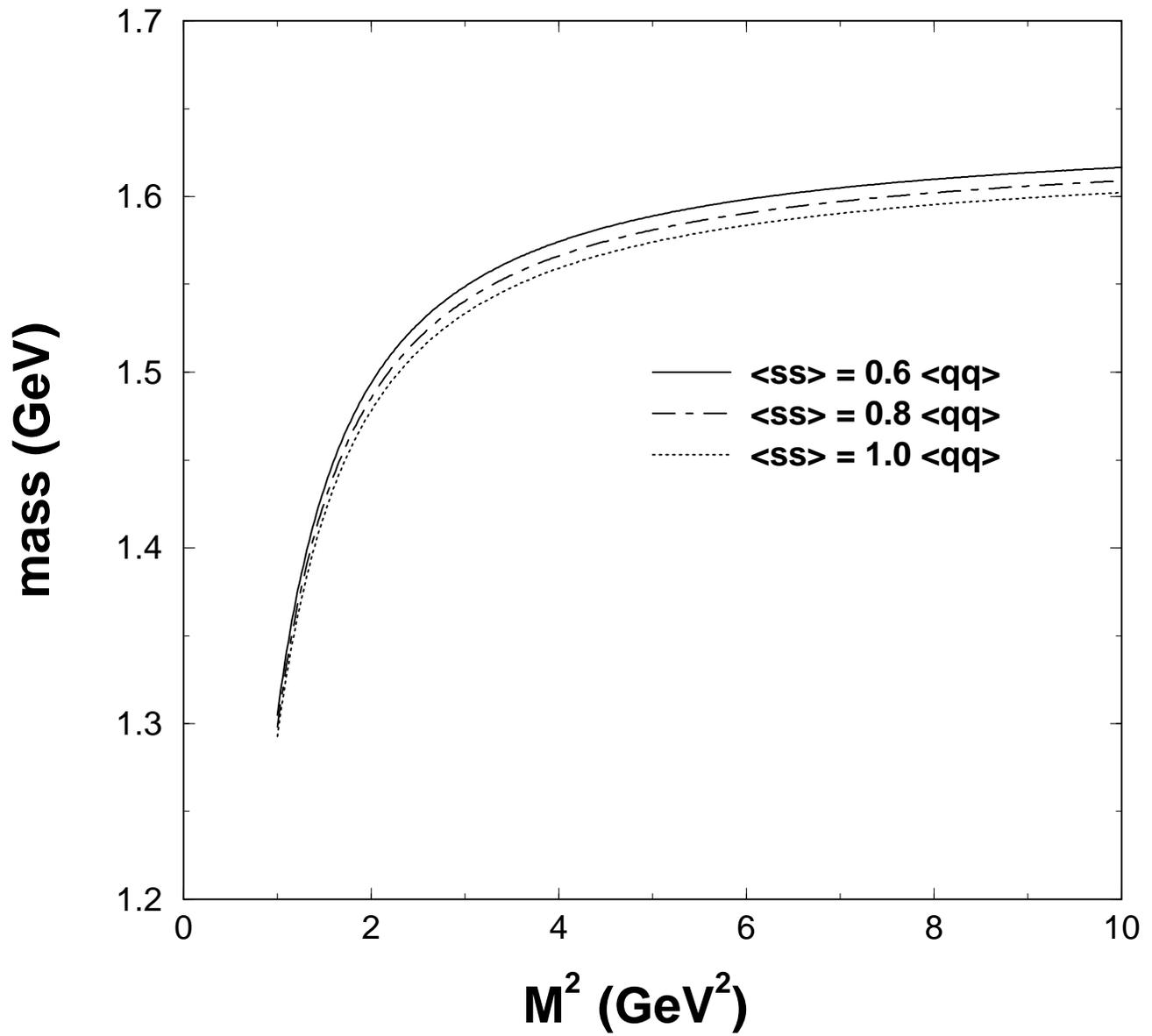
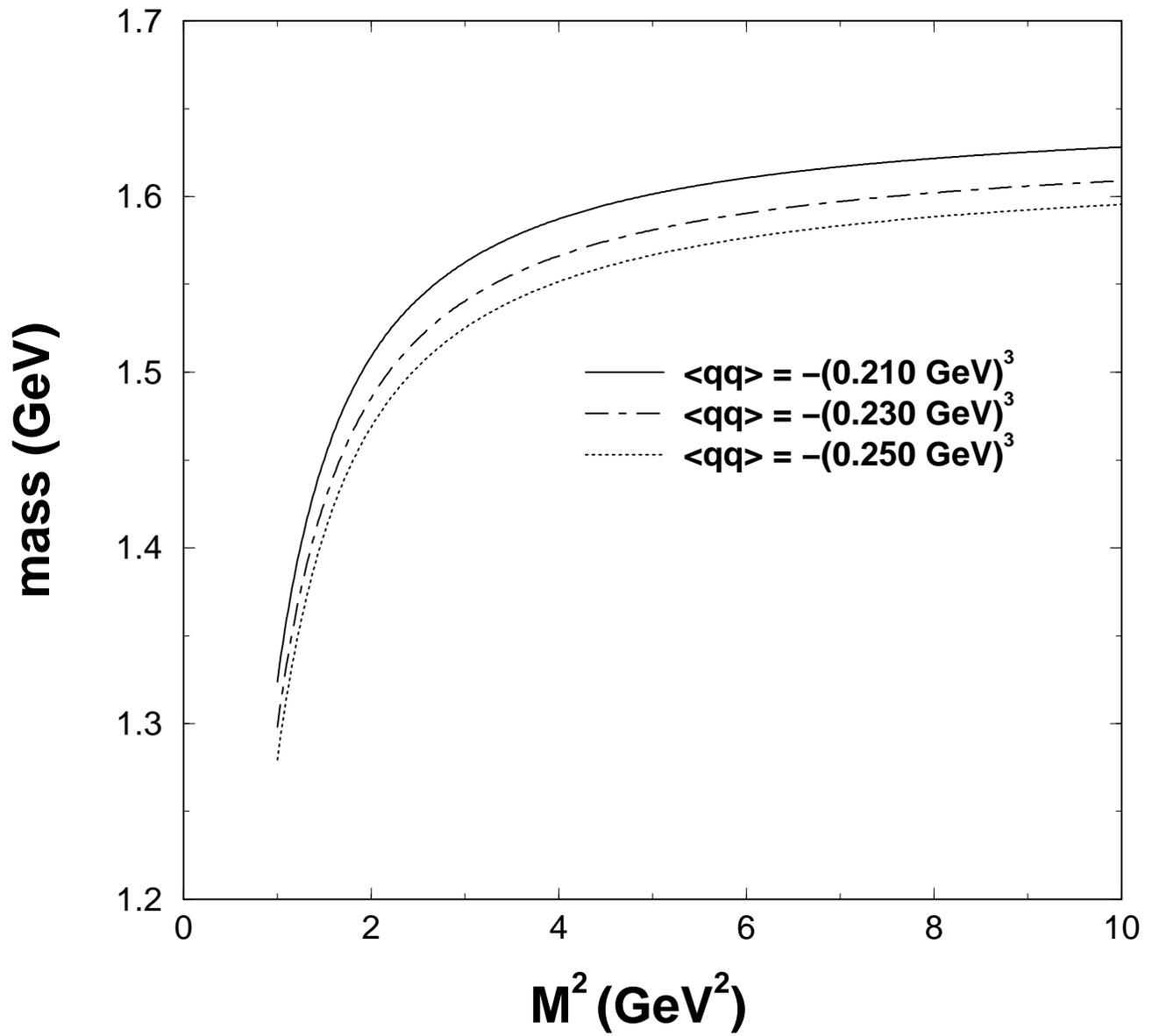
\begin{figure}
\caption{The Borel-mass dependence of the 
$\pi\Lambda$ multiquark mass at $s_0$ = 3.852 GeV$^2$
on (a) the strange quark mass (b) the strange quark condensate 
(c) the quark condensate.}
\label{fig6}
\end{figure}

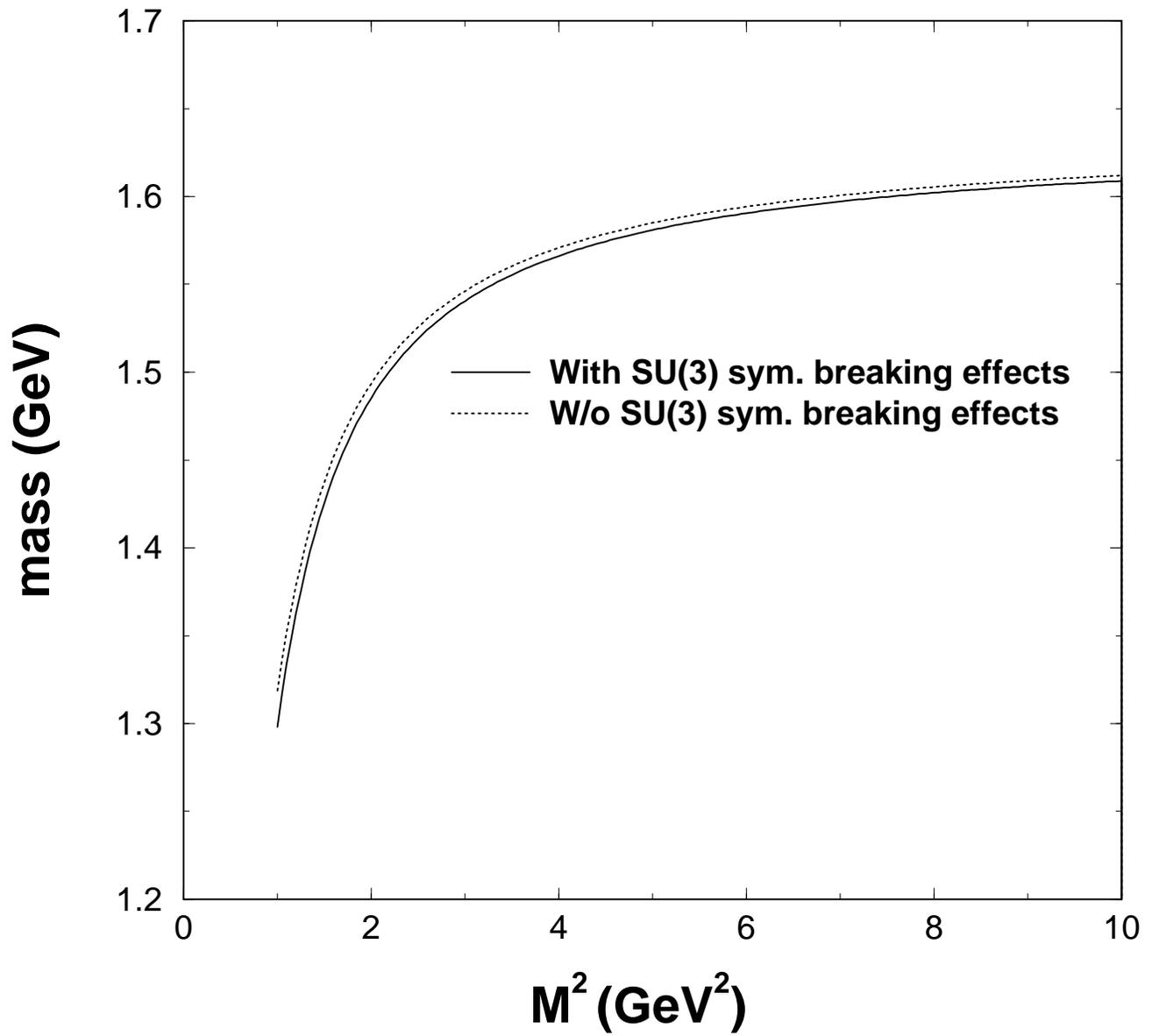
\begin{figure}
\caption{The SU(3) symmetry breaking effects in the $\pi\Lambda$
multiquark sum rule at $s_0$ = 3.852 GeV$^2$.}
\label{fig7}
\end{figure}

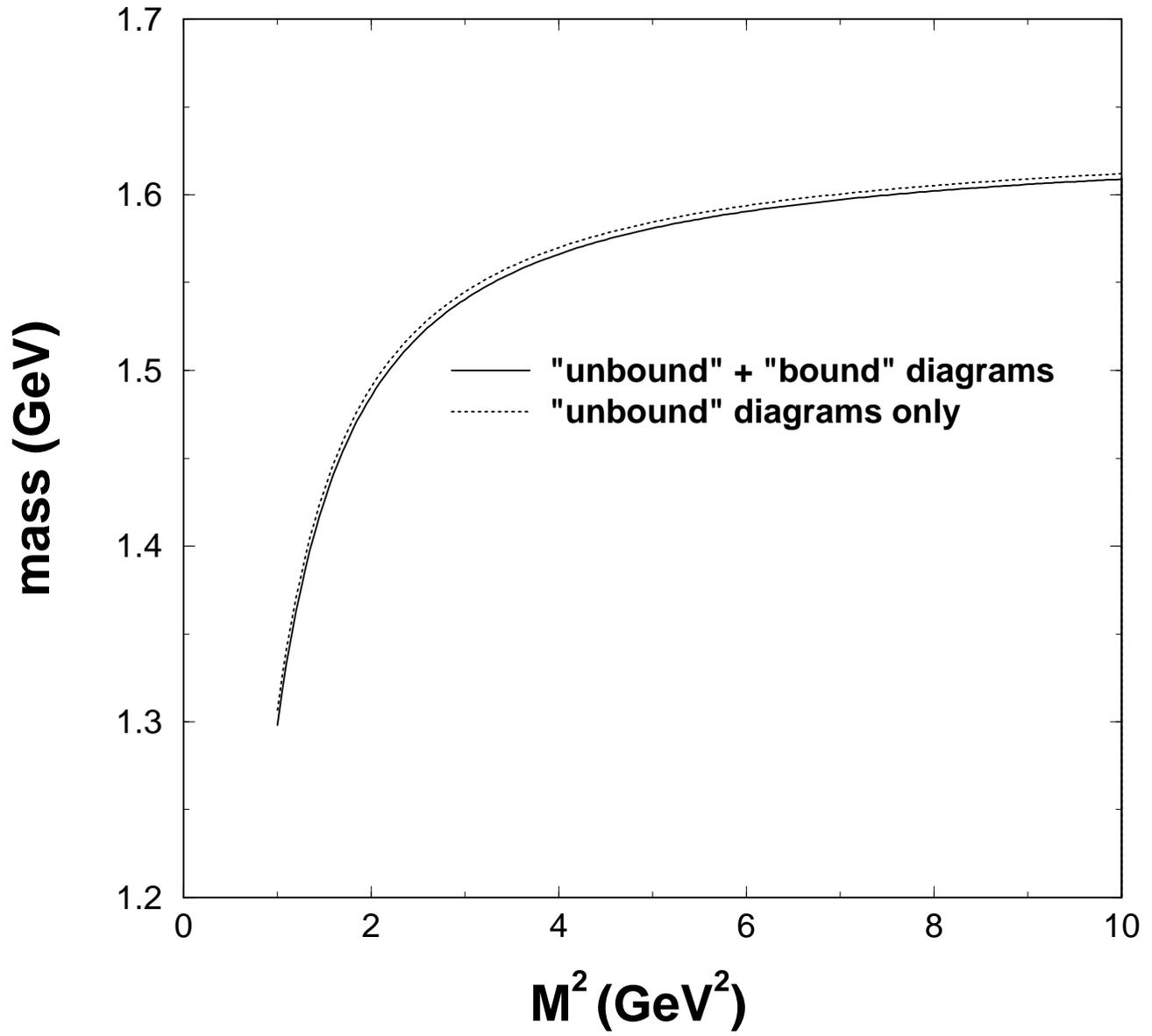
\begin{figure}
\caption{The contribution of ``bound" diagrams in the $\pi\Lambda$
multiquark sum rule at $s_0$ = 3.852 GeV$^2$.}
\label{fig8}
\end{figure}

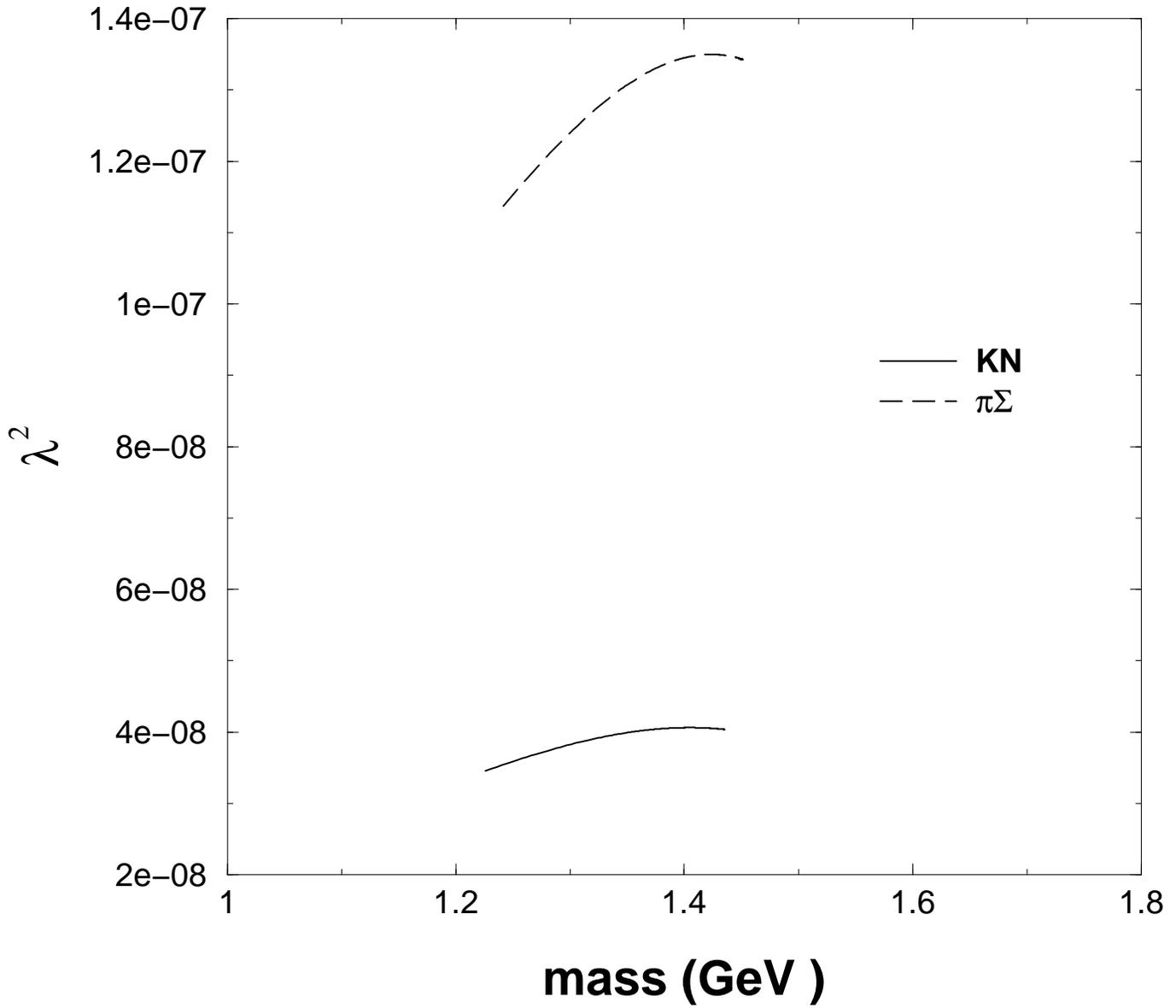
\begin{figure}
\caption{The coupling strengths and masses from the $\bar{K}N$
and $\pi^+\Sigma^- + \pi^0\Sigma^0 + \pi^-\Sigma^+$ multiquark states
at $s_0$ = 3.082 GeV$^2$.}
\label{fig9}
\end{figure}

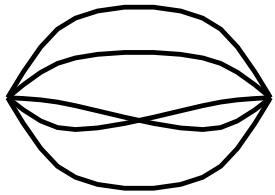
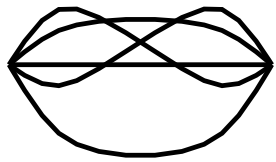
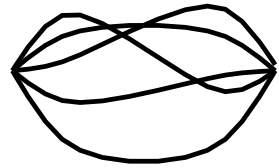
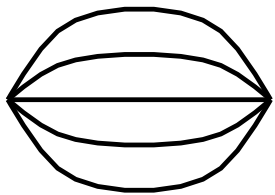
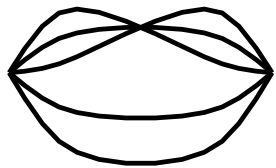
\end{document}